\documentclass{amsart}
\usepackage{cite}
\usepackage[dvips]{graphicx}

\let\leq\leqslant

\let\geq\geqslant

\usepackage{amsmath,
amsthm,
latexsym,
amssymb,
amscd,
amsgen,
amstext,
amsbsy,
amsopn,
math rsfs,
bm,
bbm,
amsthm,
epsfig,
graphicx,
graphics,
xspace,
amsxtra,
xcolor,
dsfont,
enumerate,
mathabx}

\usepackage[hidelinks]{hyperref}

\theoremstyle{remark}
\newtheorem{remark}{Remark}[section]

\newcommand{\R}{\mathbb{R}}
\newcommand{\I}{\mathcal{I}}
\newcommand{\J}{\mathcal{J}}
\newcommand{\ep}{\varepsilon}
\newcommand{\la}{\lambda}
\newcommand{\diff}{\mathrm{d}}

\newcommand{\tx}{\textstyle}
\newcommand{\lf}{\left}
\newcommand{\ri}{\right}
\newcommand{\braket}[2]{\lf\langle #1|#2 \ri\rangle}
\renewcommand{\leq}{\leqslant}
\renewcommand{\geq}{\geqslant}

\numberwithin{equation}{section}

\newcommand{\beq}{\begin{equation}}
\newcommand{\eeq}{\end{equation}}

\newcommand{\bmln}[1]{\begin{multline*} #1 \end{multline*}}

\begin{document}

\title[2D Schr\"odinger equation with nonlinear point interactions]{An introduction to the twodimensional Schr\"odinger equation with nonlinear point interactions}

\author[R.~Carlone,\,M.~Correggi,\,L. ~Tentarelli ]{$^{1}$R.~Carlone,\,$^{2}$\,M.~Correggi,$^{2}$\,L. ~Tentarelli  }
\address{$^{1}$ Universit\`{a} ``Federico II'' di Napoli, \\Dipartimento di Matematica e Applicazioni ``R. Caccioppoli''\\ MSA, via Cinthia, I-80126, Napoli, Italy\\
$^{2}$ ``Sapienza'' Universit\`{a} di Roma, \\ Dipartimento di Matematica\\ P.le Aldo Moro, 5, 00185, Roma, Italy}
\email{raffaele.carlone@unina.it, michele.correggi@gmail.com,tentarelli@mat.uniroma1.it}

\begin{abstract}
We present an introduction to the nonlinear Schr\"odinger equation (NLSE) with concentrated nonlinearities in $\mathbb{R}^{2}$. Precisely, taking a cue from the linear problem, we sketch the main challenges and the typical difficulties that arise in the twodimensional case, and mention some recent results obtained by the authors on local and global well-posedness. 
\end{abstract}

\maketitle

\section{Introduction}

In the last twenty years the Schr\"odinger equation with point interactions has proven to be a very useful mathematical tool for modeling many interesting phenomena in several areas of theoretical physics: from foundations of quantum mechanics (e.g., \cite{ft,ccf07,cfn15,cfn}) to acoustics (e.g., \cite{cfpo}), from quantum field theory (e.g., \cite{cfp}) to spectral theory (e.g., \cite{ccf11}). In addition, linear and nonlinear point interactions can be seen as singular perturbations not only of the standard Schr\"odinger (or heat) equation. They may appear, for instance, also in the study of the Dirac equation, a model which has recently attracted some renewed attention (see e.g., \cite{aghhe,cmp,CCNP})

Linear point interactions arise as a particular, but relevant, application of the more general \emph{theory of self-adjoint extension} of symmetric operators; a theory that has gained new popularity in recent years also for the application to the study of evolution equations in non-standard domains, such as \emph{quantum graphs} (see e.g., \cite{acfn1,acfn2,acfn3,acfn4,AST0,AST1,AST2,AST3,BK,CFN,GSD,N,NPS,ST1,ST2,T}) and \emph{quantum hybrids} (see e.g., \cite{CE,cp,ES1,ES2,ES3}).

The extension to nonlinear point interactions appeared first in \cite{at} and its interest is driven by the possibility of investigating nonlinear problems in the context of \emph{solvable models} (i.e., models with an explicit solution) such as, indeed, point interactions. In dimensions d$=$1,3 these problems have been extensively analyzed and many results have been obtained, such as local and global well-posedness (see \cite{at,adft}), occurrence of blow-up solutions (see \cite{at,adft04}), and approximation by  standard NLSEs with concentrating potentials (the so-called \emph{point-like limit} -- see \cite{cfnt14,cfnt}). We also mention that in these dimensions also linear \emph{non-autonomous} models have been widely studied, mainly in relation to complete \emph{ionization} phoenomena (see e.g., \cite{CCLR,CD,CDFM,SY}).

On the contrary, to the best of our knowledge, the two-dimensional problem have failed to be understood for years, even though some of the main technical difficulties arising in this case were known (see \cite{aghhe,ccf11,CCF}). However, the problem has been finally solved by the authors in \cite{CCT}, where new features of Volterra integral operators with highly singular kernels have been established (see also \cite{CF}) in order to prove local and global well-posedness of the associated Cauchy problem. 

In this communication, starting from linear point interactions in $\mathbb{R}^{2}$, we sketch some important points of the strategy used in \cite{CCT} to discuss the issues of local and global well-posedness of the nonlinear model.

\section{Linear Point Interactions}

In this section we give a brief overview of linear point interactions in $\mathbb{R}^{2}$ (for more details we refer to \cite{aghhe}).

As in $\R^3$ (and in contrast to what occurs in $\R$), in $\mathbb{R}^{2}$ the starting point is that of giving a precise meaning to the formal operator
\begin{equation}
\label{formal}
H^{f}:=- \Delta +\sum_{i=1}^{N}\alpha_{i} \delta(\cdot-\mathbf{y}_i),\qquad \mathbf{y}_{1},\dots,\mathbf{y}_{N}\in \mathbb{R}^{2},\quad\alpha_1,\dots,\alpha_N\in \mathbb{R}.
\end{equation}
Hence, we look for a suitable self-adjoint operator in $L^{2}\left(\mathbb{R}^{2}\right)$, which correctly represent the heuristic expression in \eqref{formal}. In particular, this operator has to act as the free Laplacian far from the points were the interactions are located.

In the following we recall how to construct such an operator by means of the theory of self-adjoint extensions. For the sake of simplicity, we consider only the case of a single point interaction placed in $\mathbf{y}\in \mathbb{R}^{2}$ (for the generalization to the case of a finite number of point interactions see \cite{aghhe}).

\subsection{Definition and setting}

Let us introduce the following restriction of the Laplacian:
\[
H^{r}:=-\Delta, \qquad \mathcal{D}(H^r):=C_{0}^{\infty}\left(\mathbb{R}^{2}\backslash\{\mathbf{y}\}\right).
\]
This operator is symmetric and, furthermore, acts exactly as the standard Laplacian far from the interaction point. Hence, the strategy to find a suitable form for \eqref{formal} is that of classifying all possible (non trivial) self-adjoint extensions of $H^r$.

In fact, one can prove (see \cite{aghhe}) that all these extensions are given by a one-parameter family of operators $\displaystyle H_{\alpha,\mathbf{y}}$  with domain and action given by
\[
\begin{split}
\mathcal{D}(H_{\alpha,\mathbf{y}}):=&\left\{ \psi\in L^{2}(\mathbb{R}^{2})|\,\psi=\phi_{\lambda}+q\,G^{\lambda}(\cdot-\mathbf{y}), \, \phi_{\lambda}\in H^{2}\left(\mathbb{R}^{2}\right), \right.\\[.1cm]
&\left.q\in \mathbb{C},\,\lim_{\mathbf{x}\to\mathbf{y}}\phi_\lambda(\mathbf{x})=\left(\alpha+\tfrac{1}{2\pi}\log\sqrt{\lambda}+\tfrac{\gamma}{2\pi}\right)q\right\}
\end{split}
\]
(for any $\lambda>0$) and
\[
 (H_{\alpha,\mathbf{y}}+\lambda)\psi:=(-\Delta+\lambda)\phi_\lambda,\qquad\forall\psi\in\mathcal{D}(H_{\alpha,\mathbf{y}}),
\]
where $\gamma$ is the Euler-Mascheroni constant and $G^{\lambda}$ is the Green's function of $-\Delta+\lambda$, namely $G^{\lambda}(\mathbf{x})=\frac{1}{2\pi}K_0(\sqrt{\lambda}|\mathbf{x}|)$, with $K_0(\sqrt{\lambda}|\cdot|)$ denoting the inverse (unitary) Fourier transform of $ (|\mathbf{k}|^2 + \lambda)^{-1} $, i.e., the modified Bessel function of second kind of order $ 0 $ (a.k.a. Macdonald function \cite[Sect. 9.6]{AS}).

\begin{remark}
In the 3d case an analogous construction holds, but with a major difference: one can define an equivalent decomposition (up to modifying the integrability requirements at infinity) for $\lambda=0$. Here, on the contrary, although the operator domain is independent (as in 3d) of the parameter $\lambda>0$, the choice $\lambda=0$ is forbidden due to the infrared singularity of the 2d Green's function (which in fact diverges when $ \lambda \to 0 $).
\end{remark}

\begin{remark}
The parameter $\alpha$ introduced above is not the inverse scattering lenght. There is a relation between $\alpha$  defined above with the scattering lenght $(4\pi\alpha^{scatt})^{-1}$ as defined in \cite{aghhe} and it is the following
\[
\alpha^{scatt}+\frac{\gamma}{2\pi}-\frac{\log 2}{2\pi}=\alpha.
\]
\end{remark}

The quadratic form associated with $H_{\alpha,\mathbf{y}}$, in addition, is given by:
\begin{equation}
\label{form}
\begin{split}
\mathcal{F}_{\alpha,\mathbf{y}}(\psi):=&\left\|\nabla\phi_{\lambda}\right\|_{L^{2}(\mathbb{R}^{2})}^2+\lambda\left\|\phi_{\lambda}\right\|_{L^{2}(\mathbb{R}^{2})}^2-\lambda\left\|\psi\right\|_{L^{2}(\mathbb{R}^{2})}^2+\\[.2cm]
&+\lf( \alpha + \tx\frac{1}{2\pi} \log \frac{\sqrt{\la}}{2} + \frac{\gamma}{2\pi} \ri) |q|^2,
\end{split}
\end{equation}
with form domain
\[
V:=\left\{\psi\in L^{2}(\mathbb{R}^{2})|\,\psi=\phi_{\lambda}+q\,G^{\lambda}(\cdot-\mathbf{y}),  \phi_{\lambda}\in H^{1}\left(\mathbb{R}^{2}\right), q\in \mathbb{C}\right\}.
\]
Such a quadratic form is not positive definite. As a consequence, one finds that a bound state occurs for any value of $\alpha$ (another major difference with the 3d case, where the sign of the coupling constant $\alpha$ distinguishes between the existence/nonexistence of a bound state). More precisely:
\[
\sigma \left( H_{\alpha,\mathbf{y}}\right)=\left\{-4 e^{-4\pi\alpha-2\gamma}\right\}\cup[0,\infty),
\]
and thus twodimensional point interactions can be said to be always attractive.

\subsection{Dynamics}
\label{dynamics}

As a consequence of the previous considerations (by means, for instance, of the Stone's theorem), it is well-known that, for any $\psi_{0} \in D(H_{\alpha,y})$, the Cauchy problem  
\begin{equation}
\label{cp}
\left\{\begin{array}{l}
 \displaystyle \imath\frac{\partial \psi_{t}}{\partial t}=H_{\alpha,\mathbf{y}}\psi_{t}\\[.3cm]
 \displaystyle \psi_{t=0}=\psi_0
\end{array}\right.
\end{equation}
is globally well-posed. In addition, in this case, an expression for the propagator as an integral kernel is available (see again \cite{abd,aghhe}). This means that the solution of the Cauchy problem can be given explicitly.

\medskip
However, there exists an alternative description for the dynamics of \eqref{cp}, which has two main advantages. It makes the state description more similar to the physics intuition of what a point interaction is, and, especially, it is a suitable starting point for the generalization from linear problems to nonlinear problems, where no theory of self-adjoint operators is available.

This description is based on the following ansatz for the solutions:
\begin{equation}\label{ans}
\psi_t(\mathbf{x})=(U_{0}(t)\psi_{0})(\mathbf{x})+\frac{\imath}{2\pi} \int_{0}^{t}\diff s \: U_{0}(t-s,|\mathbf{x}-\mathbf{y}|)q(s)
\end{equation}
where $U_{0}(t)$ is the integral kernel of the 2d free Schr\"odinger propagator, i.e.,
\[
U_{0}(t;\mathbf{x})=\frac{e^{-\frac{|\mathbf{x}|^{2}}{4\imath t}}}{2\imath t},
\]
and $q(t)$ is a complex scalar function usually called \emph{charge}. In this way, all the relevant information on the interaction is stored in fact in $q(t)$ and hence the dynamics of the problem is completely determined by the equation satisfied by $q(t)$, that is the so-called the \emph{charge equation}.

\begin{remark}
At an intuitive level, \eqref{ans} is simply the Duhamel formulation of \eqref{cp} if one assumes that $q(t)$ represent the dynamics of the wave function at the interaction point.
\end{remark}

Before justifying the previous ansatz on $\psi_t$ and deriving, at least formally, the charge equation, we need to introduce a technical tool (see also \cite{CCT,CF,skm}). Recall that the \emph{Volterra functions} (see, e.g., \cite{E}) are defined as
\[
\nu(t,\alpha):=\int_{0}^{\infty}\diff s \: \frac{t^{\alpha+s}}{\Gamma(\alpha+s+1)}
\]
where $\displaystyle \Gamma(t):=\int_{0}^{\infty}\diff x \: x^{t-1}e^{-x}$. In particular, we focus on the Volterra function of order $-1$, i.e., $\nu(t,-1)=:\I(t)$. This function is finite (and analytic) for every $t>0$, whereas
\[
 \I(t) \sim \frac{1}{t \log^2 \left(\frac{1}{t}\right)}\left[1 + \mathcal{O}(\left|\log t \right|^{-1}) \right],\qquad\text{as}\quad t\downarrow0,
\]
so that $\I\in L_{\mathrm{loc}}^1([0,\infty))$ and $\I\not\in L_{\mathrm{loc}}^p([0,\infty))$, for every $p>1$. In addition,
\[
 \I(t)\sim e^{t}+\mathcal{O}(t^{-1} ),\qquad\text{as}\quad t\to+\infty.
\]
Furthermore, and above all, such a function is a \emph{Sonine} kernel, namely there exists another function $\J(t)$ such that
\[
 \int_0^t \diff s\: \I(t-s)\J(s)=1\qquad\forall t\geq0.
\]
Precisely, $\J(t):=-\gamma-\log t$.

\medskip
Now, we can explain why \eqref{ans} solves \eqref{cp}, provided that $\psi_t\in\mathcal{D}(H_{\alpha,\mathbf{y}})$ for every $t\geq0$. For the sake of simplicity we assume here $q(0) = 0$, since this is not restrictive (the argument if $q(0)\neq0$ is analogous, up to further computations). First we note that
\begin{align*}
 \imath \partial_t \psi_t (\mathbf{x}) = & \, ( - \Delta U_0(t) \psi_0 ) (\mathbf{x}) - \frac{q(t)}{2\pi}	+ \frac{1}{2\pi}\int_0^t \diff\tau \: \partial_\tau U_0 (t - \tau; |\mathbf{x} - \mathbf{y}|) q(\tau)\\
                                = & \, ( - \Delta U_0(t) \psi_0 ) (\mathbf{x}) - \frac{1}{2\pi}\int_0^t \diff \tau \: U_0 (t - \tau; |\mathbf{x} - \mathbf{y}|)\dot q(\tau),
\end{align*}
where we used the fact that $i\partial_tU_0(t)\psi_0=-\Delta U_0(t)\psi_0$. Hence, applying the Fourier transform on $ \R^2 $, the above expression reads (setting $k=|\mathbf{k}|$)
\begin{equation}
 \label{eq:t_derivative_fourier}
 \imath \widehat{\partial_t\psi}_t(\mathbf{k}) =  k^2 e^{-\imath k^2 t} \widehat{\psi_0} (\mathbf{k})  - \frac{1}{2\pi} \int_0^t \diff\tau \: e^{- \imath \mathbf{k} \cdot \mathbf{y}} \,e^{-\imath k^2(t- \tau)} \dot q(\tau).
\end{equation}
On the other hand, the Fourier transform of $ H_{\alpha,\mathbf{y}}\psi_t $ turns out to be 
\bmln{
  	k^2 \bigg(\widehat{\psi}_t(\mathbf{k}) - \frac{1}{2\pi} \frac{q(t) e^{-\imath \mathbf{k} \cdot \mathbf{y}}}{k^2 + \lambda} \bigg) - \frac{\lambda}{2\pi} \frac{q(t) e^{-\imath \mathbf{k} \cdot \mathbf{y}}}{k^2 + \lambda}\\	
  	\displaystyle = k^2  e^{-\imath k^2 t} \widehat{\psi}_0 (\mathbf{k})  + \frac{1}{2\pi} \int_0^t \diff\tau \: e^{- \imath \mathbf{k} \cdot \mathbf{y}} \,\partial_\tau \left( e^{-\imath k^2(t- \tau)} \right) q(\tau) - \frac{q(t)e^{-\imath \mathbf{k}\cdot\mathbf{y}}}{2\pi}\\	
  	\displaystyle =   k^2 e^{-\imath k^2 t} \widehat{\psi}_0 (\mathbf{k})  - \frac{1}{2\pi}  \int_0^t \diff \tau \: e^{- \imath \mathbf{k} \cdot \mathbf{y}} \,e^{-\imath k^2(t- \tau)} \dot q(\tau),
}
which is equal to the r.h.s. of \eqref{eq:t_derivative_fourier}. Summing up, if $\psi_t\in\mathcal{D}(H_{\alpha,\mathbf{y}})$ for every $t\geq0$, then the ansatz \eqref{ans} solves \eqref{cp}. In fact, some regularity for the charge $q$ is also required in order to make rigorous the previous computation. However, since in view of \eqref{ans}, the regularity of $\psi_t$ is due to $q(t)$, this request is somehow hidden in the assumption $\psi_t\in\mathcal{D}(H_{\alpha,\mathbf{y}})$.

At this point it is evident that the central question is the behavior of $q(t)$, or in other words, the detection of the proper evolution equation for $q(t)$ (which will turn out to be a Volterra integral equation of the first kind). The argument below is just a formal derivation of the charge equation (a more rigorous way that exploits Laplace transform can be found in \cite{A}), but is interesting since stresses the underlying link between this equation and the boundary condition present in the operator domain.

In order to guarantee that $\psi_t\in\mathcal{D}(H_{\alpha,\mathbf{y}})$, a boundary condition must be satisfied, i.e.,
\[
 	\phi_{\lambda,t}(\mathbf{y})=\frac{1}{2\pi} \int_{\R^2} \diff\mathbf{k} \: e^{\imath \mathbf{k} \cdot \mathbf{y}} \,\widehat{\phi}_{\lambda,t}(\mathbf{k}) =  \left(\alpha+ \tfrac{1}{2\pi} \log \tfrac{\sqrt{\lambda}}{2} - \tfrac{\gamma}{2\pi} \right) q(t).
\]
Moreover, since $\phi_{\lambda,t}=\psi_t-q(t)G^{\lambda}(\cdot-\mathbf{y}),$ 
\bmln{
  	\displaystyle \frac{1}{2\pi} \int_{\R^2} \diff\mathbf{k}\: e^{\imath \mathbf{k} \cdot \mathbf{y}} \bigg\{ e^{-\imath k^2 t} \widehat{\psi}_0 (\mathbf{k}) + \frac{\imath}{2\pi} \int_0^t \diff\tau \: e^{- \imath \mathbf{k} \cdot \mathbf{y}}\, e^{-\imath k^2(t- \tau)} q(\tau)- \frac{1}{2\pi} \frac{q(t) e^{-\imath \mathbf{k} \cdot \mathbf{y}}}{k^2 + \lambda} \bigg\}\\
  	\displaystyle  = \left(\alpha + \tfrac{1}{2\pi} \log \tfrac{\sqrt{\lambda}}{2} - \tfrac{\gamma}{2\pi} \right) q(t).
}
Combining the last diverging term on the l.h.s. with the second one via an integration by parts, we get
\bmln{
  	\displaystyle \frac{1}{2\pi} \int_{\R^2} \diff\mathbf{k} \bigg\{ e^{\imath \mathbf{k} \cdot \mathbf{y}}\, e^{-\imath k^2 t} \widehat{\psi}_0 (\mathbf{k}) - \frac{1}{2\pi (k^2+\lambda)} \int_0^t \diff\tau \: e^{-\imath k^2(t- \tau)}\left[ \dot q(\tau) - \imath \lambda q(\tau) \right] \bigg\} \\
 	\displaystyle  = \left(\alpha + \tfrac{1}{2\pi} \log \tfrac{\sqrt{\lambda}}{2} - \tfrac{\gamma}{2\pi} \right) q(t).
}
The integral in $\mathbf{k}$ of the second term on the l.h.s. contains an infrared singularity for $ t = \tau $ which is proportional to $ \log(t - \tau) $: in fact by \cite[Eqs. 3.722.1 $\&$ 3.722.3]{GR}
\bmln{
 \int_{\R^2} \diff\mathbf{k} \: \frac{e^{-\imath k^{2}(t-\tau)}}{k^{2}+\lambda}  = - \pi e^{\imath \lambda(t-\tau)} \left[ \mathrm{ci}(\lambda(t-\tau)) -\imath\, \mathrm{si}(\lambda(t-\tau))  \right] \\
                                                                 = - \pi e^{\imath \lambda (t - \tau)} \left( \gamma + \log \lambda + \log (t - \tau) \right) + e^{\imath \lambda (t - \tau)} Q(\lambda; t-\tau),
}
where $\textrm{si}( \: \cdot \:)$ and $ \mathrm{ci}( \: \cdot \:) $ stand for the sine and cosine integral functions (see \cite[Eqs. 5.2.1 $\&$ 5.2.2]{AS} for the definition) and, by \cite[Eq. 5.2.16]{AS},
\[
 Q(\lambda; t - \tau) : = - \pi \bigg( \sum_{n=1}^{\infty}\frac{(-(t-\tau)^{2} \lambda^{2})^{n}}{2n(2n)!} - \imath \, \textrm{si}((t - \tau) \lambda) \bigg)
\]
(note that $ Q(0; t - \tau) = -\tfrac{\imath\pi^2}{2} $). Hence, we obtain that
\bmln{
	 \left(U_0(t) \psi_0\right)(\mathbf{y}) - \left( \alpha + \tfrac{1}{2\pi} \log \tfrac{\sqrt{\lambda}}{2} + \tfrac{\gamma}{2\pi} \right) q(t)   \\
 	= - \frac{1}{4 \pi} \int_0^t  \diff\tau \: \left( \gamma +\log (t - \tau) + \log\lambda  - \tfrac{1}{\pi} Q(\lambda; t-\tau) \right)  \partial_{\tau} \left( e^{\imath \lambda (t - \tau)} q(\tau) \right) 
}
and taking the formal limit $ \lambda \to 0 $ (notice the exact cancellation of the diverging $ \log \lambda $ terms)
\[
\begin{array}{l}
 \displaystyle (U_0(t) \psi_0)(\mathbf{y}) - \left(\alpha - \tfrac{1}{2\pi} \log 2 + \tfrac{\gamma}{2\pi} -\tfrac{\imath}{8} \right) q(t)\\
 \displaystyle \hspace{4cm}= - \frac{1}{4 \pi} \int_0^t \diff\tau \left( \gamma +\log (t-\tau)    \right)  \dot q(\tau).
 \end{array}
\]
Finally, applying the convolution integral operator defined by $\I$ and using the Sonine property, suitably rearranging terms, one has
\begin{equation}
\label{gensel}
q(t) + \int_0^t \diff\tau \: \I(t - \tau)(4 \pi \alpha-2\log 2 +2 \gamma -\tfrac{i\pi}{2})q(\tau) = f(t),
\end{equation}
where 
\begin{equation}
\label{forz}
f(t)=4\pi\int_0^t \diff\tau \: \I(t-\tau)(U_0(\tau)\psi_0)(\mathbf{y}),
\end{equation}
which is what is usually called charge equation.

\medskip
It is clear that the previous computations are just formal. Moreover, the actual strategy to prove that \eqref{ans} solves \eqref{cp} is the converse of what we made. Indeed, the main steps should be the following:
\begin{itemize}
 \item[(i)] proving that \eqref{gensel} has a unique solution, at least on a small interval;
 \item[(ii)] proving that it also displays such a regularity that $\psi_t\in\mathcal{D}(H_{\alpha,\mathbf{y}})$ (and thus \eqref{ans} satisfies \eqref{cp});
 \item[(iii)] proving that the solution of \eqref{gensel} (and consequently the solution of \eqref{cp}) is global in time.
\end{itemize}

However, one can immediately see that this strategy is not the most suitable in the linear case, since classical theory of self-adjoint operators provides immediately (i)--(iii). In addition, point (ii) is not easy to prove in a direct way since the integral operator
\[
 \lf( Ig \ri)(t):=\int_0^t \diff\tau \: \I(t-\tau)g(\tau),
\]
which is the main feature of the charge equation, has no regularizing properties in Sobolev spaces (due to its highly singular behavior at the origin) and this prevents to establish the suitable regularity on $\phi_{\lambda,t}$. More in detail, even if $g$ is a smooth function, $Ig$ may present a very rough behavior. For instance, setting $g\equiv1$, one can see that $Ig(t)=\int_0^t \diff\tau \: \I(\tau)=\nu(t,0)$ is not even in $H_{\mathrm{loc}}^\theta(\R^+)$, if $\theta>1/2$, while it belongs to $H_{\mathrm{loc}}^{1/2}(\R^+)$.

On the other hand, in the nonlinear case, when the classical theory is not available, the strategy hinted before is the unique path one can follow, provided that one can manage point (ii) in spite of the singular behavior of the operator $I$.

\begin{remark}
 The lack of regularizing properties of the integral operator $I$ is the main difference between the 2d case and the 1d and 3d ones. Indeed, in odd dimension the resulting charge equation displays the $1/2$-Abel kernel $\frac{C_\beta}{t^{1-\beta}}$, $\beta\in(0,1)$, as integral kernel (in place of $\I$), which has sufficient smoothing properties (see \cite{at,gv}) to overcome the regularity issues.
\end{remark}

\section{Nonlinear Point Interactions}

As we mentioned before, the method based on the investigation of the Duhamel formula and the charge equation, although not necessary in the linear case, is the one that solely allows an easy generalization to the nonlinear problem.

Precisely, this extension is done by analogy, simply assuming that the strength of the interaction $\alpha$ depend itself on the charge in a nonlinear way (of power type), i.e.,
\beq
	\label{eq:alpha}
	\boxed{
	\alpha = \beta_0 \lf| q(t) \ri|^{2\sigma},	\qquad		\beta_0 \in \R, \sigma \in \R^+. 
}
\eeq
Then, one has to follow the strategy suggested before, namely one has to prove that the function $\psi_t$ defined by \eqref{ans} solves \eqref{cp} at least in a weak sense, provided that there is a unique and sufficiently regular solution of the charge equation. It is clear that, in view of \eqref{eq:alpha}, \eqref{gensel} reads
\begin{equation}
 \label{eq:charge_eq}
 \begin{array}{l}
 	\displaystyle q(t)+4\pi\beta_0\int_0^t \diff\tau \: \I(t-\tau)|q(\tau)|^{2\sigma}q(\tau) \\
	\hspace{3.5cm} \displaystyle -2 \left( \log 2 - \gamma +\tfrac{\imath\pi}{4}\right)\int_0^t \diff\tau \: \I(t-\tau)q(\tau)= f(t),
 \end{array}
 \end{equation}
 where $f(t)$ is given again  by \eqref{forz}.
 
This problem has been solved in \cite{CCT} by the authors. However, since an exhaustive presentation of the proof would require the management of several hard and subtle technical issues, here we just give some hints on the strategy used to overcome items (i)--(iii).

\subsection{Sketch of the strategy}

First, we want to point out that in \cite{CCT} we dealt with the weak solution of \eqref{cp}, namely with  a function $\psi_t\in V$ that satisfies
\begin{equation}
\label{eq:cauchyweak}
\left\{\begin{array}{l}
\displaystyle \imath \frac{\diff}{\diff t} \braket{\chi}{\psi_t} =   \mathcal{F}_{\alpha,\mathbf{y}}\big[\chi,\psi_t\big]_{\big|{\lf\{ \alpha = \beta_0 \lf| q(t) \ri|^{2\sigma}\ri\}}},\\
\displaystyle \psi_{t = 0}  =  \psi_0,
\end{array}\right.
\end{equation}
for any $ \chi=\chi_\lambda+q_\chi G^{\lambda}(\cdot-\mathbf{y}) \in V $, where $\braket{\cdot}{\cdot}$ is the inner product of $L^2(\R^2)$ and 
\bmln{
\mathcal{F}_{\alpha,\mathbf{y}}\big[\chi,\psi_t\big]_{\big|{\lf\{ \alpha = \beta_0 \lf| q(t) \ri|^{2\sigma}\ri\}}} := \int_{\R^2}\diff \mathbf{x} \: \left\{\nabla\chi_\lambda^*\cdot\nabla\phi_{\lambda,t}+\lambda\chi_\lambda^*\phi_{\lambda,t}-\lambda\chi^*\psi_t\right\} \\ +\bigg(\beta_0|q(t)|^{2\sigma}+\frac{1}{2\pi}\log\frac{\sqrt{\lambda}}{2}+\frac{\gamma}{2\pi}\bigg)q_\chi^*q(t).
}
The search for a strong solution seems to be out of reach at the moment. We will explain the reason below.

\begin{remark}
 We stress that $\mathcal{F}_{\alpha,\mathbf{y}}\big[\cdot,\cdot\big]_{\big|{\lf\{ \alpha = \beta_0 \lf| q(t) \ri|^{2\sigma}\ri\}}}$ is nothing but the nonlinear analogous of the sesquilinear form associated to the quadratic form $\mathcal{F}_{\alpha,\mathbf{y}}$ defined by \eqref{form} and hence is the natural choice for the definition of weak solution.
\end{remark}

Point (i) is the simplest one since it exploits some well-known results on nonlinear Volterra integral equations (see, e.g., \cite{M,ms}). Thus, one almost immediately finds that \eqref{eq:charge_eq} has a unique continuous solution $q(t)$ on a maximal existence interval $[0,T_*)$, where $T_*$ is possibly infinite.

On the other hand, the central and more delicate point is (ii). In particular, one can prove that, in order to have that $\psi_t\in V$ and satisfies \eqref{eq:cauchyweak} in the maximal existence time, it is sufficient that the $q\in H^{1/2}(0,T)$ for all $T<T_*$. However, as we suggested in the previous section, this is not an easy task due to the lack or regularizing properties of the integral operator $I$, defined by the Volterra function $\I$. In addition, since as we showed before the action of the operator $I$ can destroy the regularity even of smooth functions, the strategy of the 1d and 3d cases, where we recall that the integral kernel $\I$ is replaced by the $1/2$-Abel one, namely solving smoother problems with more regular initial data and then using a density argument, is forbidden too.

Consequently, the unique available strategy, which is the one exploited in \cite{CCT}, is that of
\begin{itemize}
 \item developing a contraction argument on a possibly small interval $[0,T]$;
 \item repeating the same argument on consecutive intervals with suitable modifications of \eqref{eq:charge_eq};
 \item proving that the attachments preserve the regularity at the connection points and allow to cover any closed and bounded interval strictly contained in $[0,T_*)$.
\end{itemize}
Such a procedure works since the operator $I$ displays the following \emph{contractive} property (see \cite{CCT,CF} for the proof):
\begin{equation}
\label{eq:est_I_lim}
\lf\| Ig \ri\|_{H^{1/2}(0,T)} \leq C_T \left( \lf\| g \ri\|_{L^\infty(0,T)}+ \lf\| g \ri\|_{H^{1/2}(0,T)}\right),
\end{equation}
where $C_T\to0$, as $T\to0$.

Finally, the proof of point (iii) consists of detecting sufficient conditions in order to claim that $T_*=+\infty$. The first step in this direction is the proof of the the conservation of the mass, i.e., $M(t):=\|\psi_t\|_{L^2(\R^2)}$, and, especially, of the energy, i.e.,
\[
E(t): = \|\phi_{1,t}\|_{H^1(\R^2)}^2 +\left( \frac{\beta_0}{\sigma+1}|q(t)|^{2\sigma}+ \frac{\gamma-\log 2}{2\pi}\right)	|q(t)|^2.
\]
Hence, a classical blow-up alternative analysis shows that in the so-called \emph{defocusing} case, i.e., $\beta_0>0$, the solution is global in time, whereas in the \emph{focusing} case, i.e., $\beta_0<0$, $T_*$ may be both finite and infinite, depending on the initial datum $\psi_0$.

\subsection{Further remarks} The methods mentioned before to manage points (ii) and (iii) have proved to be full of subtle and hard technical issues. An extensive discussion of these goes beyond the aims of this proceeding and has been done in detail in \cite{CCT}. However there are points that deserve some comments.

First, we want to stress an immediate reason that makes the proof the strong version of \eqref{eq:cauchyweak} out of reach at the moment. This is again connected to point (ii) and, precisely, to the contracting properties of $I$. It is in fact possible to establish an analogous of \eqref{eq:est_I_lim} also for $H^1$-functions (actually for any $\nu\in(1/2,1]$), which is the regularity required to get $\psi_t\in\mathcal{D}(H_{\alpha,\mathbf{y}})$ up to the proof of the boundary condition. However, in this case it is necessary  to assume that $q(0)=0$, which is an unnatural assumption and, in addition, prevents the possibility of using the attachments technique highlighted in the previous section. Thus, point (ii) cannot be proved for the $H^1$ regularity and this prevents at the moment the possibility of finding strong solutions of \eqref{eq:cauchyweak}.

Furthermore, both the attachments technique and the proof of the energy conservation call for a further regularity of the charge. The former issue is due to the failure of the Hardy inequality for $H^{1/2}$-functions (see \cite{cfnt,KP}), that prevents the attachment of two $H^{1/2}$-functions on consecutive intervals to be in $H^{1/2}$, in general. The latter, on the contrary, is due to the  integration of the derivative of the charge, which is necessary in the computations, but which have no meaning as $q$ is not absolutely continuous, in general. However, if one proves that $q$ is \emph{log-H\"older} continuous, namely, that its modulus of continuity is controlled by a logarithmic function (in place of a fractional power function), then both the issues can be overcome. Indeed, in this case the attachments can be proved to be licit and one can develop a duality pairing argument in order to bypass the problem on integrating $\dot{q}$.

Unfortunately, the proof of such a further regularity for the charge requires some extra-assumptions on the initial datum $\psi_0$. Precisely, its regular part $\phi_{\lambda,0}$ has to satisfy
\[
 (1+k)^{\ep}\,\widehat{\phi}_{\lambda,0}\in L^1(\R^2),\qquad\text{for some}\quad\ep>0,
\]
which is a restriction with respect to the natural assumptions that only state $\psi_0\in V$.

Finally, it is worth recalling that the contractive argument needed to manage point (ii) requests a slightly restrictive assumption on the power of the nonlinearity. Precisely, one must suppose that $\sigma\geq1/2$. This is due to the fact that otherwise one cannot prove Lipschitz continuity of the map $g\mapsto|g|^{2\sigma}g$ between $H^{1/2}(0,T)\cap L^\infty(0,T)$ and itself, with a constant that do not blow up as $T\to0$; namely, with a constant that do not compensate the good contractive properties of $I$.

\end{document}